\documentstyle[twocolumn,prb,aps]{revtex}
\begin{document}
\draft
\title{Bosonic and Fermionic Single-Particle States in the Haldane Approach
to Statistics for Identical Particles}
\author{Serguei B. Isakov}
\address{Medical Radiology Research Center, Obninsk, Kaluga Region 249020,
Russia}
\date{Received 12 May 1995}
\maketitle

\begin{abstract}
We give two formulations of exclusion statistics (ES) using a
{\it variable} number of bosonic or fermionic single-particle states
which depend on the number of particles in the system. Associated
bosonic and fermionic ES parameters are introduced and are
discussed for FQHE quasiparticles, anyons in the lowest
Landau level and for the Calogero-Sutherland model. In the latter case,
only one family of solutions is emphasized to be sufficient to
recover ES; appropriate families are specified for a number of
formulations of the Calogero-Sutherland model.
We extend the picture of variable number of single-particle states to
generalized ideal gases with statistical interaction between
particles of different momenta. Integral equations are derived
which determine
the momentum distribution for single-particle states and distribution of
particles over the single-particle states in the thermal equilibrium.
\end{abstract}


The idea of reduction of single-particle Hilbert space upon adding
particles into the system introduced by Haldane into
definition of statistics for identical particles \cite{H}
has recently received great attention.
This idea was originally proposed
to give an alternative definition of fractional statistics
based on a generalized exclusion principle [exclusion
statistics (ES)], with a specific
law of reduction of the Hilbert space determined by an ES
parameter. Associated formula for the dimension of the
many-particle Hilbert space was suggested \cite{H}
which was then applied to FQHE
quasiparticles to evaluate numerically \cite{JC,HXZ} and
analytically \cite{LiO} their ES parameters in simple cases.

The latter formula applied locally in the phase space was also used as a
starting point for formulation of the statistical mechanics
\cite{I-B,Wu} (see also Ref.~7). The statistical distribution for
ES turned out to coincide with that previously derived \cite{I-A} for
fractional statistics in one dimension [the latter was introduced
in the algebraic (Heisenberg) approach to quantization of identical
particles \cite{LM-H} and was argued \cite{LM-H,P} to be modeled by
systems with an inverse square interaction (the Calogero-Sutherland
models\cite{CS})]. On related studies of ES in terms of the
Calogero-Sutherland models, see Refs.~12--15.

Interpretation of anyons confined to the lowest Landau level (LLL) in
terms of ES statistics was discovered in various ways
\cite{Wu,dVO-MPL,CJ,MSh-1}. This is consistent with the interpretation
of anyons in the LLL, where their dynamics becomes effectively
one-dimensional, in terms of fractional statistics in one
dimension in the algebraic approach\cite{HLM}.

The idea of reduction of the space of single-particle states was used to
define statistical interaction between particles of different
momenta\cite{BWu}. The latter notion was also discussed in the language
of generalized single-state statistical distributions for free identical
particles \cite{I-PRL}.

Haldane has defined ES by the condition
that upon adding $\Delta N$ particles into the system (we restrict
ourselves to the case of one species of particles), the
number of single-particle states $D$ available for further
particles is altered by
\begin{equation}
\Delta D=-g \, \Delta N \;,
\label{eq:1} \end{equation}
where $g$ is the ES parameter (with $g=0$
and $g=1$ for bosons and fermions). He also suggested the formula for
the dimension of the space of many-particle states:
\begin{equation}
W = {(D+N-1)! \over N!(D-1)!} \;.
\label{eq:2} \end{equation}
If one denotes by $G$ the number of single-particle states available for
the first particle and uses the expression
\begin{equation}
D=G-g(N-1) \;,
\label{eq:3} \end{equation}
consistent with (\ref{eq:1}), then (\ref{eq:2})
interpolates between the numbers of ways of placing $N$ bosons
($g=0$) and $N$ fermions $(g=1)$ over $G$ single-particle states.

We stress that in order to derive the dimension of the space of
many-particle states starting from the dimension of the space of
single-particle states, it is necessary to know ``accommodation''
properties of single-particle states.
In the usual interpretation of Eqs.~(\ref{eq:2})--(\ref{eq:3})
as counting ways of placing $N$ particles into $G$
states, Eq.~(\ref{eq:2}) (which in fact determines implicitly
``accommodation'' properties of single-particle states) should be viewed
as independent of (\ref{eq:1}).

In this paper we give an  alternative interpretation of Eqs.~
(\ref{eq:1})--(\ref{eq:2}), in terms of usual
bosonic single-particle states. In this case, (\ref{eq:2})
is a consequence of (\ref{eq:1}).
We also give a dual description of ES which uses fermionic
single-particle states. The two formulations of the ES are somewhat
similar to the boson-based and fermion-based descriptions of anyons.
We extend these formulations to involve local statistical interaction
in the phase space as well as statistical
interaction between particles of different energies. Based on this, we
reexamine systems of anyons in the LLL, the Calogero-Sutherland systems
and interacting systems of spinless particle solvable by the
thermodynamic Bethe ansatz (TBA), refining and extending previous results
\cite{I-A,Wu,MSh-2,BWu,dVO-MPL,I-PRL} on thermodynamic equivalence of the
above systems to systems of free identical particles.

{\it Bosonic and fermionic single-particle states.} ---
Consider a system that may be described in terms of bosonic (fermionic)
single-particle states and assume that the total number of
single-particle states depends on the number of particle in the system:
$D^{b,f}=D^{b,f}(N)$. Single-particle states may thus appear or
disappear on adding particles into the system. The formula similar to
(\ref{eq:1}) then defines variation the number of
single-particle states upon adding $\Delta N$
particles into the system:
\begin{equation}
\Delta D^{b,f}(N)=-g^{b,f} \Delta N \;,
\label{eq:5}\end{equation}
where we have introduced the {\it bosonic} and {\it fermionic} ES
parameters $g^b$ and $g^f$. In place of (\ref{eq:3}), we have
\begin{equation}
D^{b,f}=G-g^{b,f} (N-1) \,,
\label{eq:6}\end{equation}
where $G$ is the number of single-particle states in the absence of
particles. The dimension of the space of many-particle states follows
from (\ref{eq:6}) to be
\begin{equation}
W^b = {(D^b+N-1)! \over N!(D^b-1)!}\;, \;\; W^f={D^f! \over N!(D^f-N)!}\,.
\label{eq:7}\end{equation}
Note that $W^b$ and $W^f$ with $D^b$ and $D^f$ from (\ref{eq:6})
coincide, resulting in the same statistical properties, if
\begin{equation}
g^b=1+g^f \,.
\label{eq:8}\end{equation}

We see that Haldane's original ES parameter corresponds to the {\it
bosonic} ES parameter in our notation.

As an example, we consider ES parameters for FQHE quasiparticles.
The ES parameters for the quasiholes (qh) and quasielectrons (qe)
above the
Laughlin states with the filling factors $\nu=\frac1m$ with $m$ odd were
found by estimations of the dimension of
many-particle space for quasiparticles in the spherical geometry
\cite{JC,HXZ,LiO}. ES parameters were identified by comparison with
(\ref{eq:2})--(\ref{eq:3}), which corresponds to the bosonic ES
parameters. The results of Refs.~2--4 then read
$g_{\rm qh}^b=\frac1m$ and $g_{\rm qe}^b=2-\frac1m$. According to
(\ref{eq:8}), we may as well say that the fermionic ES
parameters for the qh and qe are
$$
g_{\rm qe}^f=-g_{\rm qh}^f=1-1/m \;,
$$
possessing  the symmetry $g_{\rm qe}^f=-g_{\rm qh}^f$ which was
originally suggested for the bosonic ES parameters \cite{H}.

We thus have the two descriptions of ES, in terms of bosonic and
fermionic single-particle states, respectively. This is reminiscent of
the two descriptions of anyons in two spatial dimensions:
anyons with the same exchange statistics
parameter $\theta$ with $0\leq\theta <2\pi$ can be described either as
bosons carrying the magnetic flux (in units of quantum flux
$\phi_0=2\pi/e$, $e>0$) $\tilde\phi^b=\theta/\pi$  with
$0\leq\tilde\phi^b <2$ or as fermions carrying the magnetic flux
$\tilde\phi^f$ with $-1\leq\tilde\phi^f <1$, where the two descriptions
are connected by
\begin{equation}
\tilde\phi^b=1+\tilde\phi^f \;.
\label{eq:9}\end{equation}
This relation is quite similar to (\ref{eq:8}). We show below that
for anyons confined to the LLL where the notion of ES
applies, this similarity becomes the precise correspondence: the
bosonic and fermionic ES parameters for anyons in the
LLL exactly coincide with $\tilde\phi^b$ and $\tilde\phi^f$.

{\it Particles of the same energy and anyons in the LLL.} ---
Let all the particles have the same energy $\varepsilon$.
It follows from (\ref{eq:7}) that the entropy $S= \ln W$ is
\cite{TDlimit}
\begin{equation}
S^{b,f}=D^{b,f}[\pm (1 \pm n^{b,f}) \ln (1 \pm n^{b,f})
-n^{b,f}\ln n^{b,f}]
\label{eq:11} \end{equation}
(the upper and lower signs refer to the bosonic-state and
fermionic-state descriptions, respectively), where
\begin{equation}
n^{b,f}\equiv N/D^{b,f}=n/(1-g^{b,f}n) \;,
\label{eq:11a}\end{equation}
and $n=N/G$. We maximize (\ref{eq:11}) with
respect to $n^{b,f}$ subject to the constraints of fixed total
number of particles $N=D^{b,f}n^{b,f}$
and the total energy $E=D^{b,f}\varepsilon n^{b,f}$, with
the associated Lagrange multipliers $\beta\mu$ and $-\beta$
($\beta=1/T$, and $\mu$ is the chemical potential). With (\ref{eq:11a}),
we obtain the equation for $n^{b,f}$ in the thermal equilibrium:
\begin{equation}
n^{b,f}(1\pm n^{b,f})^{\pm g^{b,f}-1}=x \;,
\label{eq:12a}\end{equation}
where $x=e^{\beta (\mu -\varepsilon)}$.

The equation of state can be derived from (\ref{eq:11}) and (\ref{eq:12a})
using the relations $\Omega =E-TS-\mu N$ and
$\Omega=-PA$ (specified for two spatial dimensions with A
the area occupied by the gas) to be
\begin{equation}
P\beta = \pm (G/A)\ln(1\pm n^{b,f}) \,.
\label{eq:13} \end{equation}

Let us compare this equation of state with that obtained for anyons in
a strong magnetic field \cite{dVO-PRL}, where only particles in the LLL
(all having the same energy) contribute into the equation of state.
Regarding anyons as bosons carrying the magnetic flux of
value $\tilde \phi^b$ $(0\leq \tilde \phi^b <2)$ in the direction
antiparallel to the external magnetic field $B$, the equation of state
was found to be \cite{dVO-PRL} $P\beta = \rho_L\ln[1+\nu
/(1-{\tilde\phi}^b\nu )]$, where $\rho_L= eB/2\pi$ is the
density of states in the LLL, $\nu=N/\rho_LA$ is the filling factor.
We may as well rewrite this equation
of state regarding anyons as fermions carrying magnetic flux
${\tilde\phi}^f$ ($-1\leq {\tilde\phi}^f <1$)
connected with ${\tilde\phi}^b$ by (\ref{eq:9}). The two equations of
state coincide with (\ref{eq:13}), if one identifies the density of
states $G/A$ in (\ref{eq:13}) with $\rho_L$ (and hence, $n$ with $\nu$),
and, in addition, makes the identifications
\begin{equation}
g^{b,f}={\tilde \phi}^{b,f} \,.
\label{eq:15} \end{equation}

Let us comment on formula (\ref{eq:6}) in this context.
Consider the mean-field approximation corresponding to the change of a
flux carried by a particle with a uniform magnetic field of the same
flux. Adding a particle into the system then diminishes the total magnetic
flux by ${\tilde \phi}^{b,f}$, hence the the number of states in the LLL,
equal to the total magnetic flux of in units of the flux quantum, is
\begin{equation}
D^{b,f}=eBA/2\pi-{\tilde \phi}^{b,f}N \,.
\label{eq:14a}\end{equation}
This corresponds to (\ref{eq:6}) with $g^{b,f}$ from (\ref{eq:15}). But we
stress that (\ref{eq:14a}) holds only in the mean-field approximation.

{\it Statistical interaction between particles of the same momentum.} ---
Let now particles in different single-particle states may have different
energies. For definiteness, we assume that single-particle states may be
labeled with the momenta which particles have in these states.
We apply the construction of the previous section to particles of the
same momentum $k$ (supplying all the extensive quantities of
the previous section with the subscript $k$, which turns all the
quantities into densities per unit momentum).
We assume that upon adding particles of momentum $k$ into the
system, there may appear or disappear single-particle states of only
the same momentum:
\begin{equation}
D_k^{b,f}=G_k-g^{b,f}N_k \;.
\label{eq:6a} \end{equation}
The entropy for the whole system is\cite{G_k}
\begin{equation}
S^{b,f}=\sum_k G_k \ln W_k^{b,f} \,,
\label{eq:7a} \end{equation}
where $W_k^{b,f}$ are obtained from (\ref{eq:7}) by the changes $N\to
N_k$ and $D^{b,f}\to D_k^{b,f}$.
Again, the statistical properties given by the bosonic and fermionic
pictures are identical if the condition (\ref{eq:8}) is fulfilled.

For $n_k^{b,f}$ $(\equiv N_k/D_k^{b,f})$ in equilibrium, in place of
(\ref{eq:12a}), we have
\begin{equation}
n_k^{b,f}(1\pm n_k^{b,f})^{\pm g^{b,f}-1}=x_k \;,
\label{eq:8b}\end{equation}
where $x_k=e^{\beta (\mu -\varepsilon_k)}$, and $\varepsilon_k$ is the
energy of a particle with momentum $k$. The substitution
$n_k^{b,f}=n_k/(1-g^{b,f}n_k)$, yields the equation for $n_k$
$(\equiv N_k/G_k)$:
\begin{equation}
n_k(1-g^{b,f}n_k)^{\mp g^{b,f}}(1-g^{b,f}n_k\pm n_k)^{\pm
g^{b,f}-1}=x_k.
\label{eq:12b} \end{equation}

For the bosonic ES parameter, (\ref{eq:12b}) recovers the
equation that was derived in the formulation of
ES where (\ref{eq:2}) was viewed as counting the ways of placing
$N$ particles over $G$ states \cite{I-B,Wu}. That formulation implied {\it
invariable} number of single-particle states, and the function $n_k$
played the part of the distribution of particles over states $k$
(`statistical distribution'). Unlike that, in the
present formulation of ES, with {\it variable} number of single-particle
states, the distribution of particles over states is given by the
functions $n_k^{b,f}$.

Eq.~(\ref{eq:12b}) for the bosonic ES parameter
was discussed in the context of the Calogero-Sutherland model as a
system which reveals ES \cite{I-A,MSh-2,BWu,dVO-MPL}. Here we extend
that analysis with specific emphasis on the fact that only one family of
solutions of this model is sufficient to recover ES.

We discuss the Calogero-Sutherland model in a harmonic well governed by
the Hamiltonian
\begin{equation}
H=-{1\over 2}\sum_{i=1}^N{\partial^2 \over {\partial x_i^2}}+
\sum_{i<j} \frac{\lambda(\lambda-1)}{(x_i-x_j)^2}
+\frac12\omega^2\sum_i x_i^2 .
\label{eq:25a}\end{equation}
Consider the coordinate domain $x_1\leq \cdots \leq x_N$. Due
to the symmetry of the Hamiltonian (\ref{eq:25a}) under the change
$\lambda\to 1-\lambda$, the system (\ref{eq:25a}) has two classes of
solutions of the form
\begin{equation}
\psi^{\rm I}(\lambda)=\Delta^{\lambda}\Phi_+(\{x_l\},\lambda),\;
\psi^{\rm II}(\lambda)=\Delta^{1-\lambda}\Phi_+(\{x_l\},1-\lambda),
\label{eq:26a}\end{equation}
where $\Delta=\prod_{i>j}(x_i-x_j)$, $\{x_l\}$ is the set with
$l=1,\dots,N$, and $ \Phi_+(\{x_l\},\lambda)=P(\{x_l\},\lambda)\exp
(-\frac12\sum_i x_i^2) $, where $P(\{x_l\},\lambda)$ is a
polynomial symmetric in the particle coordinates $\{x_l\}$ with a
non-zero free term. The solutions $\psi^{\rm I}(\lambda)$ and $\psi^{\rm
II}(\lambda)$ are quadratically integrable and non-singular for
$\lambda\geq 0$ and for $\lambda\leq 1$, respectively.

We may rewrite the solutions of class II as
$\psi^{\rm II}(\lambda)=\Delta^{-\lambda}\Phi_-(\{x_l\},-\lambda)$, where
$\Phi_-(\{x_l\},-\lambda)=\Delta\Phi_+(\{x_l\},1-\lambda)$ is an
antisymmetric function of the coordinates.
We can then extend continuously the solutions $\psi^{\rm I}$ for
$\lambda\geq 0$ and $\psi^{\rm II}$ for $\lambda\leq 1$ to the regions
$x_{Q1}\leq \cdots \leq x_{QN}$, where $Q$ is a permutation
of $1,\dots,N$, in the symmetric and antisymmetric way, respectively.
Making, in addition, the changes $\lambda\to\lambda_B$ in $\psi^{\rm I}$ and
$-\lambda\to\lambda_F$ in $\psi^{\rm II}$, we arrive at symmetric
(`boson') and antisymmetric (`fermion') wave functions $\psi^B$ and
$\psi^F$ of the form
\begin{equation}
\psi^{B,F}=|\Delta|^{\lambda_{B,F}}\Phi_{\pm}(\{x_l\},\lambda_{B,F}) \,,
\label{eq:26b}\end{equation}
with $\lambda_B\geq 0$ and $\lambda_F\geq -1$. These function are
solutions of the boson and fermion Hamiltonians $H^B$ and $H^F$ which
are obtained from the Hamiltonian (\ref{eq:25a}) by the changes of the
two-body potential $V(x)$ with the potentials
\begin{equation}
V^{B,F}(x)=\lambda_{B,F}(\lambda_{B,F}\mp 1)/x^2 \,,
\label{eq:27a}\end{equation}
where $\lambda_B \ge 0$ and $\lambda_F\ge -1$.

The energy levels for the solutions $\psi^{\rm I}$ and $\psi^{\rm II}$
or, equivalently, the solutions $\psi^B$ and $\psi^F$ (\ref{eq:26b}),
are given by $E^{B,F}= E_0^{B,F}+\sum_{l=1}^{N}ln_l$,
where $\{n_l\}$ is a set of non-negative integers obeying
$0\le n_1 \le \cdots \le n_N$, and
$E_0^B=\omega [{1 \over 2}N+ {1 \over 2} \lambda_B N(N-1)]$ and
$E_0^F=\omega [{1 \over 2}N+ {1 \over 2} (1+\lambda_F) N(N-1)]$
are the ground state energies (see also Ref.~23). The corresponding
partition functions are ($q=e^{-\beta\omega}$)
\begin{equation}
Z_N^{B,F}= e^{- \beta E_0^{B,F}} \prod_{l=1}^N (1-q^l)^{-1} \,.
\label{eq:27} \end{equation}

The family of solutions $\psi^B$
(\ref{eq:26b}) for the boson Calogero-Sutherland model (or,
equivalently, the family of solutions of class I for (\ref{eq:25a}))
was analyzed in Ref.~8 in the limit $\omega\to 0$ in terms of
generalized single-state partition functions and statistical
distributions in the approach with
nonvariable number of single-particle states, resulting in the
statistical distributions (\ref{eq:12b}) with $g^b=\lambda_B$.

Repeating the analysis of Ref.~8 for the family of
solutions $\psi^F$ (\ref{eq:26b}) for the fermion Hamiltonian (or,
equivalently, for the solutions of class II in (\ref{eq:26a})), with the
partition functions $Z_N^F$ (\ref{eq:27}), we
arrive at the statistical distributions $n_k$ coinciding with those
given by (\ref{eq:12b}) with $g^f=\lambda_F$.

We note that the analysis given in Ref.~13 implies using the family of
the fermion solutions $\psi^F$ of (\ref{eq:26b}) with the parameter
$g$ used in Ref.~13 equal to our $\lambda_F+1$. Refs.~12 and 15
use the family $\psi^B$ (\ref{eq:26b}).

{\it Statistical interaction between particles of different momenta.} ---
We now give a generalization of the considerations of the
previous section assuming that
on adding particles of momentum $k$ into the
system, there may appear or disappear single-particle states
of any other momenta (cf. Eq.~(\ref{eq:6a})):
\begin{equation}
D_k^{b,f}=G_k-\sum_{k'} G_{k'} g^{b,f}(k,k') N_{k'} \,,
\label{eq:21} \end{equation}
where $g^{b,f}(k,k')$ may be called the bosonic and fermionic
statistical interaction functions.

The entropy (\ref{eq:7a}) is written as
\begin{eqnarray}
S^{b,f}&=&\sum_k G_k
[\pm (d_k^{b,f}\pm n_k)\ln (d_k^{b,f}\pm n_k)   \nonumber \\
&& \mp d_k^{b,f}\ln d_k^{b,f}-n_k\ln n_k] \, ,
\label{eq:24} \end{eqnarray}
where
\begin{equation}
d_k^{b,f}\equiv D_k^{b,f}/G_k =1-\sum_{k'}G_{k'} g^{b,f}(k,k')n_{k'} \,.
\label{eq:24c}\end{equation}
It follows from (\ref{eq:24})--(\ref{eq:24c}) that the bosonic and fermionic
pictures result in the same statistical properties if\cite{1D}
\begin{equation}
g^b(k,k')=\frac{2\pi}{L}\delta(k-k')+g^f(k,k')
\label{eq:23} \end{equation}
generalizing (\ref{eq:8}).

Maximization of the entropy with respect to $n_k$,
subject to the total particle number and total energy constraints,
yields the equation for $n_k$ in equilibrium
\begin{eqnarray}
\lefteqn{
\ln \frac{n_k}{x_k}=\sum_{k'}G_{k'} \left\{
\pm g^{b,f}(k',k)\ln d_{k'}^{b,f}  \right. }  \nonumber \\
&& \left.  + \left[
\mbox{$\frac{2\pi}{L}$}
\delta(k'-k)\mp g^{b,f}(k',k) \right]
\ln ( d_{k'}^{b,f}\pm n_{k'} )  \right\} ,
\label{eq:24b}\end{eqnarray}
with $d_k^{b,f}$ of (\ref{eq:24c}).
Eq. (\ref{eq:24b}) can be rewritten as an equation for the distribution
of particles over single-particle states $n_k^{b,f}$:
\begin{equation}
\ln \frac{n_k^{b,f}}{x_k}=\sum_{k'}G_{k'}
\left[\mbox{$\frac{2\pi}{L}$}\delta(k'-k)\mp g^{b,f}(k',k) \right]
\ln (1\pm n_{k'}^{b,f} ) .
\label{eq:25b}\end{equation}
Eq.~(\ref{eq:25b}) together with the equation for the distribution of
single-particle states $d_k^{b,f}$,
\begin{equation}
d_k^{b,f}=1-\sum_{k'}G_{k'}g^{b,f}(k,k')n_{k'}^{b,f}d_{k'}^{b,f} \,,
\label{eq:25c}\end{equation}
which follows from (\ref{eq:24c}) and the equality
$n_k=d_k^{b,f}n_k^{b,f}$, enable one in principle to
determine all the thermodynamic quantities of the ideal gas (e.\ g.\
$E=\sum_k G_kd_k^{b,f}\varepsilon_kn_k^{b,f}$).
For $g^{b,f}(k,k')\propto \delta(k-k')$,
Eqs.~(\ref{eq:24b}) and (\ref{eq:25b}) reduce to Eqs.~(\ref{eq:12b}) and
(\ref{eq:8b}), respectively.

The fermionic state description is convenient to study low-temperature
properties since $n_k^f$ has a non-singular simple form for $T=0$.
Indeed, for $T=0$, all the states with $|k|\leq k_0$, where $k_0$ is
some boundary momentum (the `Fermi momentum'), are
occupied, the others are empty, that is, $n_k^f=1$ for $|k|\leq k_0$,
and $n_k^f=0$ for $|k|> k_0$. From Eq.~(\ref{eq:25c})
we then get for $T=0$ the integral equation for the
distribution of the fermionic states
\begin{equation}
d_k^f=1-\sum_{|k'|\leq k_0}G_{k'}g^f(k,k')d_{k'}^f \, ,
\label{eq:27b}\end{equation}
as well as the equation for $n_k$ ( $-k_0\leq k\leq k_0$)
\begin{equation}
n_k=1-\sum_{|k'|\leq k_0}G_{k'}g^f(k,k')n_{k'} \, .
\label{eq:28b}\end{equation}
For sufficiently low temperatures, $n_k^f$ will slightly differ from
$n_k^f|_{T=0}$.

Refs.~19 and 14  argued that the TBA equations for spinless
particles may be viewed as encoding statistical interaction between
particles of different momenta. We now discuss relation of the
above single-particle state definition of statistics to state counting
in the TBA.

In the description of the TBA we follow Refs.~25 and 26. Let
the wave function have the Bethe ansatz form
\begin{equation}
\Psi(x_1, \dots,x_N) = \sum_P A(P) \exp (i\sum_{j=1}^N k_{Pj} x_j)
\label{eq:31}\end{equation}
in the asymptotic region $x_1 \ll\cdots \ll x_N$, where
$k_1 >\cdots > k_N$, and $P$ is a permutation of $1, 2 \dots N$.
The coefficients $A(P)\equiv A(k_{P1},\dots ,k_{PN})$
are related by $ A(\dots k',k\dots)/A(\dots k,k' \dots)=S(k-k')$, where
$S(k-k')=-e^{-i\theta(k-k')}$ is the two-body `scattering matrix', and
$\theta(k-k')$ is the two-body `phase shift', antisymmetric in
$k-k'$, with $k$ and $k'$ the particle momenta.
Imposing periodic boundary conditions on the wave function (\ref{eq:31})
yields the Bethe ansatz equations $ e^{ik_jL}\prod_{l\neq j}
S(k_j-k_l)=1$ for allowed momentum values.
Note that the wave function (\ref{eq:31}), which vanishes
if $k_j=k_l$ for any two momenta ($S(0)=-1$),
implies description of the system in terms of fermionic states.

In the thermodynamic limit the Bethe ansatz equations results in the
following picture. The density of levels $r^f$ (being
the sum of the particle and hole densities, $r^f=\rho +\rho_h$) over
which the particles are distributed is determined by the integral
relation\cite{TBAnorm}
\begin{equation}
2\pi r^f(k)=1-\int\varphi(k-k')\rho (k')L\,dk' \,,
\label{eq:35}\end{equation}
with $\varphi (k)= \frac1{L}\partial\theta(k)/\partial k$ the derivative
of the phase shift, thus depending on the particle density.
Fermionic nature of the states is also displayed in the expression for
the entropy:
$$
S^f=\int [-(r^f-\rho)\ln (r^f-\rho) +r^f\ln r^f-\rho\ln \rho ]L\,dk .
$$

We see that the TBA picture exactly corresponds to the
above definition of statistics in terms of fermionic states:
Eq.~(\ref{eq:35}) and the entropy coincide with (\ref{eq:24c}) and
(\ref{eq:24}) after the identifications $n_k \leftrightarrow 2\pi\rho$,
$d_k^f \leftrightarrow 2\pi r_k^f$,
and
\begin{equation}
g^f(k,k')=\varphi(k-k') \,.
\label{eq:37} \end{equation}

Ref.~14 used the same fermionic state counting in the TBA as above
but the bosonic statistical interaction function. With this
remark, the identification (\ref{eq:37}) agrees with that obtained in
Ref.~14 if one takes into account the relation (\ref{eq:23}). But we stress
that interpretation of the formula of reduction of the space of
single-particle states (\ref{eq:21}) in the TBA scheme of Refs.~25 and 26
necessarily implies that single-particle states are to be
specified as fermionic. On the other hand, it is natural to expect that
in an alternative, bosonic state counting in the TBA, which is discussed
in Ref.~28, the reduction formula~(\ref{eq:21}) with the bosonic
single-particle states should be used.

As the first example, we consider the system of bosons with the two-body
$\delta$-function potential $V(x)=c\delta(x)$. That this system models
a possible statistics for identical particles in one dimension was
pointed out long ago \cite{LM-1}. The two-particle phase shift reads
\cite{YY}
$\theta(k)=-2\tan^{-1}(k/c)$. Eq.~(\ref{eq:37}) then yields
$g^f(k,k')=-2c/L[c^2+(k-k')^2]$. In the limits $c\to 0$ and $c\to\infty$
we have $g^f(k,k')=\frac{2\pi}{L}\delta (k-k')$ and $g^f(k,k')=0$,
recovering the thermodynamics for free bosons and free fermions as it
should be \cite{YY}.
Note that after the identifications  $n_k \leftrightarrow 2\pi\rho$ and
(\ref{eq:37}), Eq.~(\ref{eq:28b}) coincides with the $T=0$ equation for
the momentum distribution for bosons
with the $\delta$-function interaction\cite{LL}.

Consider now the Calogero-Sutherland system (\ref{eq:25a})
(without the harmonic potential).
The Schr\"odinger equation for the relative problem of two particles
$(-\partial_x^2+\lambda(\lambda-1)/x^2)\psi=k^2\psi$, where
$x=x_2-x_1$ and $k=\frac12(k_1-k_2)$ are the relative
coordinate and momentum, reduces to a Bessel equation so that any
solution of the Schr\"odinger equation can be represented as
a linear combination of the two solutions, $\sqrt{x}J_{\lambda-1/2}(kx)$ and
$\sqrt{x}J_{1/2-\lambda}(kx)$, if $\lambda-\frac12$ is a noninteger number
(for $\lambda-\frac12$ integer, the two above solutions are linear
dependent, and another fundamental system of
solutions, $\sqrt{x}J_{\lambda-1/2}(kx)$ and $\sqrt{x}Y_{\lambda-1/2}(kx)$,
involving Bessel functions of the first and second kinds, should be
used). These two solutions, for $\lambda \geq 0$ and
$\lambda\leq 1$, respectively, correspond to the solutions of classes I and
II for the Calogero-Sutherland model discussed in the previous section.

Consider the solution $\sqrt{x}J_{\lambda-1/2}(kx)$ for
$\lambda \geq 0$. For $kx\gg 1$, it has the
asymptotics $\pi^{-1/2} \cos (kx-\frac12 \pi\lambda)$. Comparing this
with the asymptotic wave function (\ref{eq:31}) for
two particles, $\Psi\propto e^{-ikx}+S(k)e^{ikx}$, we get
$S(k)=e^{-i\pi\lambda}$ for $k>0$. In order to determine the phase shift
$\theta(k)=i\ln(-S(k))$ for all the values of $\lambda$ simultaneously,
we consider the $\ln z$ on its Riemann surface glued of the complex planes
$D_k$ ($k=0,\pm 1,\pm 2, \dots$) which are
cut along their positive semiaxes and are
characterized by $2k\pi<\arg z<2(k+1)\pi$; in the $D_0$, the branch
of the logarithm $\ln z=\ln|z|+i\arg z$ with $0<\arg z<2\pi$ is chosen.
Then we obtain $\theta(k)=\pi(\lambda-1)$
for $k>0$, and accounting for the antisymmetry of the phase shift,
finally, $\theta(k)=\pi(\lambda-1)\,{\rm sgn}(k)$. This
recovers the result obtained by Sutherland\cite{S}, our choice of the
unique solution of the scattering problem thus corresponds to that of
Ref.~26.

Eq.~(\ref{eq:37}) then yields
$g^f(k,k')=\frac{2\pi}{L} (\lambda_B-1)\delta(k-k')$
According to (\ref{eq:23}), the same statistics may be described by the
bosonic statistical interaction function
$g^b(k,k')=\frac{2\pi}{L}\lambda_B\delta(k-k')$.

Similar considerations show that the fermion
Calogero-Sutherland system is equivalent to a free system with
$g^f(k,k')=\frac{2\pi}{L}\lambda_F\delta(k-k')$.
These results agree with those obtained in the previous section
for the Calogero-Sutherland system in a harmonic well.

In conclusion, we have proposed a formulation of the Haldane approach
to statistics for spinless identical particles considering a {\it
variable} number of bosonic (fermionic) single-particle states which
depends on the number of particles in the system. The variation of the
number of single-particle states is governed by the bosonic
(fermionic) statistical interaction function which reduces to
a single parameter for exclusion statistics [bosonic (fermionic) ES
parameter]. Thermodynamic quantities for the ideal gas are determined
by the momentum distributions for single-particle states together with
distributions of particles over single-particle states. The equations
for these two distributions in the most generic case are
Eqs.~(\ref{eq:25b})--(\ref{eq:25c}). The fermionic state picture seems
to be the most convenient for studying the low-temperature properties.

The picture of variable number of single-particle state exactly
corresponds to the description of states in the TBA for spinless
particles. We also note that the proposed
formulation of ES allows one to avoid encountering negative probabilities
which arise in the formulation of ES within the approach with
nonvariable number of single-particle states \cite{NW,P-pre}.

Generalization to several species of particles is possible and will be
reported elsewhere. We also expect that the above formulation of
exclusion and more general statistics can be extended to spinning
particles where many-particle states of more general permutation
symmetry than totally symmetric or antisymmetric ones should be used.

I thank J.~M. Leinaas, J. Myrheim, and S. Ouvry for helpful discussions.
I also acknowledge NORDITA and Department of Physics of University of
Oslo for their hospitality while this work was in progress. The work
was supported in part by the Russian Foundation for Fundamental
Research, grant No. 95-02-04337.


\begin{references}
\bibitem{TDlimit} Hereafter, for brevity, we write the formulas in the
 thermodynamic limit neglecting the quantities which are small compared
 to the large numbers like $N$, $G$, and $D$.
\bibitem{G_k} For one dimension, with $L$ the length
occupied by the gas, transition from summation to integration
is made by the changes $G_k\to\frac1{2\pi}L$ and $\sum_k\to\int dk$.
\bibitem{1D} Hereafter the formulas are specified for the case of one
dimension.
\bibitem{TBAnorm} We retain, for $\rho$ and $r_k^f$, the normalization
 used in the TBA for which $N=\int \rho L dk$.
\end{references}
\end{document}